\documentclass[aps]{revtex4}
\begin{document}
\title{Collective excitations in a fermion-fermion mixture with
different Fermi surfaces}
\author{Lan Yin}
\email{yinlan@pku.edu.cn}
\affiliation{School of Physics, Peking
University, Beijing 100871, P. R. China}
\date{\today}
\begin{abstract}
In this paper, collective excitations in a homogeneous
fermion-fermion mixture with different Fermi surfaces are studied.
In the Fermi liquid phase, the zero-sound velocity is found to be
larger than the largest Fermi velocity.  With attractive
interactions, the superfluid phase appears below a critical
temperature, and the phase mode is the low-energy collective
excitation.  The velocity of the phase mode is proportional to the
geometric mean of the two Fermi velocities. The difference between
the two velocities may serve as a tool to detect the superfluid
phase.
\end{abstract}
\maketitle
\section{Introduction}
A lot of progress has been made in the area of low temperature
Fermi gases in recent several years. Particularly, the quantum
degenerate regime was reached \cite{DeMarco} and Feshbach
resonance was observed \cite{Loftus}. Currently, a lot of effort
are concentrated on creating and studying the superfluid phase
\cite{OHara}.

At low temperature, $s$-wave scattering is the dominant
interaction between atoms.  Interactions with higher angular
momentum are ineffective for cooling.  To take advantage of the
$s$-wave interaction, the fermion system must contain more than
one species.  In most experiments so far, fermions are trapped in
two different hyperfine-spin states. The two Fermi surfaces are
usually different, which makes the system more complex.

Collective excitations are important properties of low-temperature
systems.  Landau predicated the existence of the zero sound in
Fermi liquids. In $^3$He systems, sound modes are distinct
signatures of exotic pairing phases.  The collective excitations
in Fermi gases have been extensively studied
theoretically\cite{Yip,Vichi,Baranov,BruunPRL,Bruun}. However, so
far, most studies assume the two species have the same Fermi
surface, which is often not true in experiments.

In this paper, we study the collective excitations in the
fermion-fermion mixture with different Fermi surfaces. For
simplicity, we consider a homogeneous system. We find that the
dispersion of collective excitations are affected by the Fermi
velocities. The superfluid phase can be exclusively identified by
its sound velocity as proposed in Ref.\cite{Baranov,BruunPRL}.

The Hamiltonian describing the fermion-fermion mixture is given by
\begin{equation}
H={\hbar^2 \over 2m_a} \nabla \psi_a^\dagger \cdot \nabla \psi_a
+{\hbar^2 \over 2m_b} \nabla \psi_b^\dagger \cdot \nabla \psi_b+ g
\psi_a^\dagger \psi_b^\dagger \psi_b \psi_a,
\end{equation}
where $m_a$ and $m_b$ are the masses of the $a$- and $b$-species.
The single-particle dispersion is given by
 $\epsilon_{\bf{k}}^{a,b}={\hbar^2 k^2 \over 2 m_{a,b}}-
\mu_{a,b}$, where $\mu_{a}$ and $\mu_{b}$ are the chemical
potentials of the two species. The coupling constant between the
two species is given by $g$.  In this paper, only the s-wave
scattering is considered, and the interaction between atoms of the
same species is ignored.

To obtain the spectrum of collective excitations, we construct the
kinetic equations, $$i \hbar {\partial n \over \partial
t}=[n,H],$$ where $n$ is a 2 by 2 density matrix given by
\begin{equation}
n_{{\bf k} {\bf k}'}=\left( \begin{array}{cc}
a_{\bf k}^\dagger a_{{\bf k}'} & a_{\bf k}^\dagger b_{-{\bf k}'}\\
 b_{-{\bf k}} a_{{\bf k}'} & b_{-{\bf k}} b_{-{\bf k}'}^\dagger
\end{array} \right).
\end{equation}
In this paper, we consider the low-energy and low-temperature
region, where the system is in the collisionless region and the
collision integral can be ignored. The density fluctuation $\delta
n$ obeys the simple kinetic equation
\begin{equation} \label{cleq}
\omega \delta n_{{\bf k} {\bf k}+{\bf q}}=\delta n_{{\bf k} {\bf
k}+{\bf q}} \epsilon^0_{{\bf k}+{\bf q}}-\epsilon^0_{\bf k} \delta
n_{{\bf k} {\bf k}+{\bf q}}+ n_{\bf k}^0 \delta \epsilon_{{\bf k}
{\bf k}+{\bf q}}-\delta \epsilon_{{\bf k} {\bf k}+{\bf q}} n_{{\bf
k}+{\bf q}}^0,
\end{equation}
where $n^0_{\bf k}$ is the density in equilibrium,
$\epsilon^0_{\bf k}$ is the mean-field energy, $\delta \epsilon$
is the energy fluctuation, and they are all 2 by 2 matrices.

In the low-frequency and long-wavelength limit, we apply gradient
expansion to second order, and eq.(\ref{cleq}) is reduced to
\begin{equation}\label{keq}
\omega \delta n_{\bf k}=[\delta n_{\bf k}, \epsilon_{\bf k}^0]+\{
\delta n_{\bf k}, {1 \over 2} {\bf q} \cdot \nabla_{\bf k}
\epsilon_{\bf k}^0 \} +[n_{\bf k}^0,\delta \epsilon_{\bf k}]-\{
\delta \epsilon_{\bf k}, {1 \over 2} {\bf q} \cdot \nabla_{\bf k}
n_{\bf k}^0 \}+[{1 \over 8} q_i q_j {\partial^2 n_{\bf k}^0 \over
\partial_{k_i} \partial_{k_j}}, \delta \epsilon_{\bf k}],
\end{equation}
where the second derivative of $\epsilon^0_{\bf k}$ can be ignored
because the fermi energies are usually much bigger than the
pairing gap. The fluctuations in energy and density are also
related through interaction, $$\delta \epsilon_{\bf
k}^{(1),(2),(3)}=g \int {d^3 k' \over (2 \pi)^3} \delta n_{{\bf
k}'}^{(1),(2),(3)},$$ $$\delta \epsilon_{\bf k}^{(0)}=-g \int {d^3
k' \over (2 \pi)^3} \delta n_{{\bf k}'}^{(0)},$$ where $\delta
\epsilon=\delta \epsilon^{(i)} \sigma_i$, $\delta n=\delta n^{(i)}
\sigma_i$, $\sigma_i$'s are Pauli matrices, and $\sigma_0$ is the
identity matrix.  As a consequence of the $s$-wave scattering,
$\delta \epsilon_{\bf k}$ is independent of ${\bf k}$. So in the
following, we omit its subscript ${\bf k}$.

\section{Zero Sound}
In the Fermi-liquid phase, the low-energy collective excitations
come from density fluctuation. Since the pairing fluctuation is
massive, the off-diagonal matrix elements $\epsilon^{(1),(2)}$ and
$n^{(1),(2)}$ can be ignored.  The mean-field energy and density
are given by
$$\epsilon^0_{\bf k}=\left(\begin{array}{cc}\epsilon_{\bf k}^a & 0 \\
0 & -\epsilon_{\bf k}^b\end{array} \right),$$
$$n^0_{\bf k}=\left(\begin{array}{cc} f(\epsilon_{\bf k}^a) & 0 \\
0 & 1-f(\epsilon_{\bf k}^b)\end{array} \right),$$ where
$f(\epsilon)=1/(1+e^{\beta \epsilon})$ is the fermi function.

At low temperature, $\nabla_{\bf k} f(\epsilon_{\bf k}^{a,b})
\approx \delta(\epsilon_{\bf k}^{a,b}) v_F^{a,b} \hat{k}$, the
kinetic equations are approximately given by
\begin{equation}
(\omega-v_F^{a,b} \hat{k} \cdot {\bf q}) \delta n_{\bf
k}^{a,b}=v_F^{a,b} \hat{k} \cdot {\bf q} \delta(\epsilon_{\bf
k}^{a,b})\delta \epsilon^{a,b},
\end{equation}
where $\delta \epsilon^{a,b}=\delta \epsilon^{(3)} \pm \delta
\epsilon^{(0)}$, $\delta n^{a,b}=\delta n^{(3)} \pm \delta
n^{(0)}$, and $v_F^{a,b}$ are the Fermi velocities. Since the
density fluctuation is always around the Fermi surfaces, it is a
good approximation to assume $\delta n_{\bf
k}^{a,b}=\delta\epsilon^{a,b} \nu_{\hat{k}}^{a,b}$, where
$\nu_{\hat{k}}^{a,b}$ are functions of $\hat{k}$, ${\bf q}$, and
$\omega$. The kinetic equations are now reduced to
\begin{eqnarray}
(\omega-v_F^{a,b} \hat{k} \cdot {\bf q})
\nu_{\hat{k}}^{a,b}=v_F^{a,b} \hat{k} \cdot {\bf q}
\delta(\epsilon_{\bf
k}^{a,b}), \\
\delta \epsilon^{a,b}=g \int{dk^3 \over (2\pi)^3}
\nu_{\hat{k}}^{b,a} \delta \epsilon^{b,a}.
\end{eqnarray}

The above equations can be further simplified to the following
form
\begin{equation}
1=g^2 D_a(x_a) D_b(x_b),
\end{equation}where $$D_{a,b}(x_{a,b})=N_{a,b}(0) \int{\hat{q} \cdot
\hat{k} d \hat{k} \over 4\pi (x_{a,b}-\hat{q} \cdot \hat{k})},$$
$x_{a,b}=\omega / (v_F^{a,b} q)$, and $N_{a,b}(0)$ are the
densities of states of the two species. The function $D_{a}(x_a)$
has an imaginary part when $x_a<1$. Therefore Eq. (\ref{zs}) has
an undamped solution only when
\begin{equation}
\omega>\max(v_F^a,v_F^b) q.
\end{equation}

In the weak coupling limit, the zero sound velocity is
approximately given by the largest Fermi velocity,
\begin{equation} \label{zs}
\omega \approx \max(v_F^a,v_F^b) q.
\end{equation}
For stronger couplings, the difference between the sound velocity
and the largest Fermi velocity becomes bigger.

\section{Phase Mode}
With attractive interactions, $g<0$, the system can go into a
pairing phase below a critical temperature.  The fermion
excitations have a gap $\Delta$ which can be obtained from the gap
equation in the standard BCS formulism,
\begin{equation}
1=-{g \over V}\sum_{\bf k} {\tanh[{\beta \over 2}(E_{\bf k}+\epsilon_{\bf k}^-)]+
\tanh[{\beta \over 2}(E_{\bf k}-\epsilon_{\bf k}^-)] \over 4 E_{\bf k}},
\end{equation}
where $E_{\bf k}=\sqrt{{\epsilon_{\bf k}^+}^2+\Delta^2}$,
$\epsilon_{\bf k}^\pm=(\epsilon_{\bf k}^a \pm\epsilon_{\bf
k}^b)/2$, and $\Delta$ is positive for simplicity.  The fermion
excitations in this phase have two branches with dispersions given
by $E_{\bf k}\pm \epsilon_{\bf k}^-$.

At zero temperature, there is no pairing contribution from the
region where $E_{\bf k}<|\epsilon_{\bf k}^-|$, as shown in the gap
equation,
\begin{equation}\label{ZT}
1=-{g \over V}\sum_{\bf k} {\theta(E_{\bf k}-|\epsilon_{\bf k}^-|)
\over 2 E_{\bf k}}.
\end{equation}
When  the two Fermi surfaces are different, there is no infra-red
divergence on the right-hand side of Eq. (\ref{ZT}) in the limit
$\Delta \rightarrow 0$. The coupling constant has to be bigger
than a critical value, $g>g_c$, for the pairing phase to be
stable. In contrast, when the two Fermi surfaces are identical,
the critical coupling constant is zero, $g_c=0$, and the pairing
of fermions is stronger.

In the following, we consider the case $g>g_c$ and at zero
temperature for simplicity. The mean-field energy and density
matrix are given by
\begin{eqnarray*}
\epsilon_{\bf k}^0=\epsilon_{\bf k}^- I+\epsilon_{\bf k}^+ \sigma_3+\Delta \sigma_1, \\
n_{\bf k}^0={1 \over 2} I-\Delta \theta_{\bf k} \sigma_1-\phi_{\bf k} \sigma_3,
\end{eqnarray*}
where $\theta_{\bf k}={1 \over 2 E_{\bf k}}$ and $\phi_{\bf
k}=\epsilon_{\bf k}^+ \theta_{\bf k}$.  The kinetic equation Eq.
(\ref{keq}) is now given by
\begin{equation} \label{eq1}
\Omega_{\bf k} \delta n_{\bf k}=M_{\bf k} \delta \epsilon,
\end{equation}
where $\Omega_{\bf k}$ and $M_{\bf k}$ are 4 by 4 matrices given
by
\begin{eqnarray*}
\Omega_{\bf k} &=& \left(\begin{array}{cccc}
\omega-\eta_{\bf k}^- & 0 & 0 &-\eta_{\bf k}^+ \\
0 & \omega-\eta_{\bf k}^- & -2i \epsilon_{\bf k}^+ & 0 \\
0 & 2i \epsilon_{\bf k}^+ & \omega-\eta_{\bf k}^- & -2i \Delta \\
-\eta_{\bf k}^+ & 0 & 2i\Delta & \omega-\eta_{\bf k}^-
\end{array} \right), \\
M_{\bf k} &=& \left(\begin{array}{cccc}
0 &\eta_{\bf k}^+ \theta_{\bf k}' \Delta & 0 & \eta_{\bf k}^+ \phi_{\bf k}' \\
\eta_{\bf k}^+ \theta_{\bf k}' \Delta & 0 &
2i \phi_{\bf k} + i {\eta_{\bf k}^+}^2 \phi_{\bf k}' & 0 \\
0 & -2i \phi_{\bf k}-i {\eta_{\bf k}^+}^2 \phi_{\bf k}' & 0 &
2i\Delta\theta_{\bf k}+{i \over 4}{\eta_{\bf k}^+}^2\Delta\theta_{\bf k}'' \\
\eta_{\bf k}^+ \phi_{\bf k}' & 0 & -2i\Delta\theta_{\bf k}-{i
\over 4}{\eta_{\bf k}^+}^2 \Delta \theta_{\bf k}'' & 0
\end{array} \right),
\end{eqnarray*}
where $\eta_{\bf k}^\pm=v_F^\pm \hat{k} \cdot {\bf q}$, $\phi_{\bf
k}'= {d \phi_{\bf k} \over d \epsilon_{\bf k}^+}$, $\phi_{\bf
k}''={d^2 \phi_{\bf k} \over {d \epsilon_{\bf k}^+}^2}$,
$\theta_{\bf k}'$ and $\theta_{\bf k}''$ are similarly defined.
The kinetic energy $\epsilon_{\bf k}^\pm$ is linearized around the
place where $\epsilon_{\bf k}^+=0$, $\nabla_{\bf k} \epsilon_{\bf
k}^\pm=v_F^\pm \hat{k}$. The curvature of $\epsilon_{\bf k}^\pm$
is negligible as long as the chemical potentials of the two
species are much larger than the pairing gap.  To a good
approximation, $v_F^\pm=(v_F^a \pm v_F^b )/2$. The fluctuations in
energy and density are related through interaction
\begin{equation} \label{eq2}
\delta \epsilon=\lambda g \int {d^3 k \over (2 \pi)^3} \delta n_{\bf k},
\end{equation}
where  $\lambda$ is the 4 by 4 matrix given by
\begin{eqnarray*}
\lambda=\left(\begin{array}{cccc} -1 & 0 & 0 & 0\\0 & 1 & 0 & 0 \\
0 & 0 & 1 & 0 \\ 0 & 0 & 0 & 1 \end{array} \right).
\end{eqnarray*}

Using Eq. (\ref{eq1}) and Eq. (\ref{eq2}), we obtain the equation
for the dispersion of the collective mode
\begin{equation}
\det |I-\lambda g \int {d^3k \over (2 \pi)^3} \Omega_{\bf k}^{-1} M_{\bf k}|=0.
\end{equation}
To the leading nontrivial order of $\omega$ and $q$, the above
equation is reduced to
\begin{equation}\label{mx}
\det \left| \begin{array}{cccc}1 & 0 & 0 & 0 \\ 0 & -g N_+(0) & 0 & 0 \\
0 & 0 & {g N_+(0) \over 4 \Delta^2}(\omega^2-{1 \over 3} v_F^a v_F^b q^2)
&{ig N_+(0) \over 2 \Delta} \omega \\0 & 0 & -{ig N_+(0) \over 2 \Delta} \omega
& 1+g N_+(0) \end{array} \right| =0,
\end{equation}
where $N_+(0)$ is the density of states of $\epsilon_{\bf k}^+$,
approximately given by $N_+(0)=2 N_a(0) N_b(0)/[N_a(0)+N_b(0)]$.
As shown in Eq.(\ref{mx}), the spin fluctuation and the pairing
amplitude fluctuation are decoupled from the the rest
fluctuations.  The uniform density fluctuation is closely coupled
to the pairing phase fluctuation.

The dispersion of the phase mode to the leading order of $q$ is
given by
\begin{equation}
\omega=q \sqrt{{1 \over 3} v_F^a v_F^b (1+g N_+(0))}.
\end{equation}
In the case of weak coupling, the phase-mode velocity is
noticeably smaller than the zero-sound velocity given by
Eq.(\ref{zs}). This difference between the two velocities is large
enough to be used in detecting the existence of the paring state,
as propose in Ref.\cite{Baranov,BruunPRL}.  Further work for the
case of trapped systems are needed to compare with experiments.

The author would like to thank T.-L. Ho and S.-K. Yip for helpful
discussions, and acknowledge support from SRF for ROCS, SEM and
support from NSFC under Grant No. 10174003.

\end{document}